\documentstyle[11pt,aaspp4]{article}

\lefthead{Calzetti et al.}
\righthead{ISO Photometry of Starbursts}

\begin{document}

\title{The Dust Content and Opacity of Actively Star-Forming 
Galaxies\altaffilmark{1}}

\author{Daniela Calzetti}
\affil{Space Telescope Science Institute, 3700 San Martin Dr., 
   Baltimore, MD 21218, USA; e-mail: calzetti@stsci.edu}
\author{Lee Armus}
\affil{SIRTF Science Center/Caltech, Pasadena, CA 91125, USA; 
e-mail: lee@ipac.caltech.edu}
\author{Ralph C. Bohlin and Anne L. Kinney}
\affil{Space Telescope Science Institute, MD 21218, USA; e-mail: 
bohlin@stsci.edu, kinney@stsci.edu}
\author{Jan Koornneef}
\affil{Kapteyn Astronomical Institute, University of Groningen, Groningen, 
9700 AV, The Netherlands; e-mail: J.Koornneef@astro.rug.nl}
\and
\author{Thaisa Storchi-Bergmann}
\affil{Instituto de Fisica, Universidade Federal Rio Grande do Sul, Porto 
Alegre, RS 91501-970, Brazil; e-mail: thaisa@if.ufrgs.br}

\altaffiltext{1}{Based on observations with ISO, an ESA project with
     instruments funded by ESA Member States (especially the PI
     countries: France, Germany, the Netherlands and the United
     Kingdom) with the participation of ISAS and NASA.}

\begin{abstract}
We present far-infrared (FIR) photometry at 150$\mu$m and 205$\mu$m of
eight low-redshift starburst galaxies obtained with the Infrared Space
Observatory (ISO) ISOPHOT. Five of the eight galaxies are detected in
both wavebands and these data are used, in conjunction with IRAS
archival photometry, to model the dust emission at
$\lambda\gtrsim$40~$\mu$m. The FIR spectral energy distributions
(SEDs) are best fitted by a combination of two modified Planck
functions, with T$\sim$40--55~K (warm dust) and T$\sim$20--23~K (cool
dust), and with a dust emissivity index $\epsilon=2$. The cool dust
can be a major contributor to the FIR emission of starburst galaxies,
representing up to 60\% of the total flux. This component is heated
not only by the general interstellar radiation field, but also by the
starburst itself. The cool dust mass is up to $\sim$150 times larger
than the warm dust mass, bringing the gas-to-dust ratios of the
starbursts in our sample close to Milky Way values, once rescaled for
the appropriate metallicity.  The ratio between the total dust FIR
emission in the range 1--1000~$\mu$m and the IRAS FIR emission in the
range 40-120~$\mu$m is $\sim$1.75, with small variations from galaxy
to galaxy. This ratio is about 40\% larger than previously inferred
from data at mm wavelengths. Although the galaxies in our sample are
generally classified as `UV-bright', for four of them the UV energy
emerging shortward of 0.2~$\mu$m is less than 15\% of the FIR
energy. On average, about 30\% of the bolometric flux is coming out in
the UV--to--nearIR wavelength range; the rest is emitted in the
FIR. Energy balance calculations show that the FIR emission predicted
by the dust reddening of the UV--to--nearIR stellar emission is within
a factor $\sim$2 of the observed value in individual galaxies and
within 20\%~ when averaged over a large sample. If our sample of local
starbursts is representative of high-redshift (z$\gtrsim$1),
UV-bright, star-forming galaxies, these galaxies' FIR emission will
be generally undetected in sub-mm surveys, unless (1) their bolometric
luminosity is comparable to or larger than that of ultraluminous FIR
galaxies and (2) their FIR SED contains a cool dust component.
\end{abstract}


\section{Introduction}

The FIR SEDs of galaxies provide insight into the energetic processes
taking place in the sources. Emission beyond a few $\mu$m from
quiescent or star-forming galaxies is dominated by dust re-radiating
the stellar energy absorbed at UV--optical wavelengths. Thus, the
description of the radiative processes in the FIR regime needs to
account for both the dust composition and the nature of the heating
sources.

Although dust emission is a complex process involving a continuum of
values for the properties of the dust grains, a simple two (or three)
components dust model has proven successful at describing the basic
characteristics of the UV--optical extinction and the FIR emission of
the dust in our own and similar galaxies (Andriesse 1978, Draine \&
Anderson 1985, Cox, Kr\"ugel \& Mezger 1986, Mathis \& Whiffen 1989,
Puget \& L\'eger 1989, Des\'ert, Boulanger \& Puget 1990, Helou, Ryter
\& Soifer 1991, Rowan-Robinson 1992). One component, sometimes split
into two (Des\'ert et al. 1990), consists of very small
grains and large molecules (size$\lesssim$100~\AA) which 
account for characteristics of the mid-IR emission, at
$\lambda\lesssim$40--50$\mu$m. This component is heated by the
single-photon absorption process to temperatures of a few hundreds
Kelvin, up to $\sim$1000~K, and is not in thermal equilibrium with its
environment (Sellgren 1984; Draine \& Anderson 1985). The second
component is formed by ``large'' grains (size$>$100~\AA) in thermal
equilibrium with their environment. These large dust grains account
for practically all the emission longward of 80--100~$\mu$m in
galaxies. Their FIR emission can be modelled as the superposition of
multiple modified Planck functions covering a range of temperatures
and modulated by the grains' emissivity
e($\nu$)$\propto\nu^{\epsilon}$, with 1$\le\epsilon\le$2 (e.g. Seki \&
Yamamoto 1980, Mezger, Mathis \& Panagia 1982, Hildebrand 1983,
Andriesse 1974, Bianchi, Davies \& Alton 1999).

The color temperatures of the large dust grains depend on both the
energy density per unit volume and the hardness of the stellar
radiation field. As a first approximation, the galaxy FIR emission
longward of 40--50~$\mu$m has been ascribed to dust heated by massive,
ionizing stars (warm dust) if the color temperature is of the order of
30--70~K and to dust heated by non-ionizing stars (cool dust) if the
temperatures are $\lesssim$15--30~K (e.g., Helou 1986, Chini, Kr\"ugel
\& Kreysa 1986a, Lonsdale-Persson \& Helou 1987, Rowan-Robinson \&
Crawford 1989, Rowan-Robinson \& Efstathiou 1993). The radiation field
created by the diffuse non-ionizing population of a galaxy is
generally referred to as the Interstellar Radiation Field (ISRF, e.g.,
Boulanger et al. 1988). Massive stars can also heat the dust to color
temperatures T$\gtrsim$70~K; this hot dust contributes, together with
the small, non-equilibrium grains mainly to the mid-IR emission
shortward of $\sim$40~$\mu$m, typically with decreasing filling factor
for increasing temperature (e.g., Natta \& Panagia 1976). The
intensity of the dust emission at $\lambda\sim$10~$\mu$m has been
observed to decrease for increasing star-formation activity in a
galaxy (Helou 1986); this has been interpreted as the destruction of
large molecules/very small grains by hard radiation from massive stars
(Boulanger et al. 1988). Throughout the paper, `thermal FIR emission'
refers to the emission in the wavelength range 1--1000~$\mu$m from
large dust grains with equilibrium temperatures
T$\lesssim$60--70~K. `Total FIR emission' defines the integrated dust
emission from both small and large grains, also in the range 1~$\mu$m
to 1000~$\mu$m.

The largest database of FIR data of galaxies available to date is from
the IRAS survey (e.g., Soifer et al. 1989). Because the wavelength
coverage is 8--120~$\mu$m, IRAS observations alone are not adequate
for characterizing the emission from dust cooler than
T$\approx$30~K. However, cool dust has been shown to be a major
contributor to the total dust mass content of a galaxy (e.g., Kwan \&
Xie 1992, Xu \& Helou 1986) and to contribute to the opacity
budget. In relatively recent years, observations at mm and sub-mm
wavelengths have been used to complement IRAS data, to address the
issue of the cool dust content of galaxies (e.g., Chini et al. 1986b,
Kwan \& Xie 1992, Roche \& Chandler 1993, Sievers et al. 1994,
Rigopoulou, Lawrence \& Rowan-Robinson 1996, Andreani \& Franceschini
1996, to quote a few). One of the drawbacks of mm observations is the
potential presence of contamination from thermal and non-thermal radio
emission (e.g., Roche \& Chandler 1993). Less sensitive to this
problem, but still effective at characterizing dust emission down to
temperatures of $\lesssim$15~K, are observations obtained with the ISO
satellite, which covered the FIR wavelength range up to
$\sim$240~$\mu$m. ISO data have shown that the FIR emission of late
type galaxies receives a substantial contribution from cool dust
(T$\sim$10--20~K, Kr\"ugel et al. 1998, Haas et al. 1998, Alton et
al. 1998, Domingue et al. 1999), increasing by a factor 2--3, up to
10, the dust content of such systems relative to previous estimates,
often based on IRAS data alone; the cool dust component is extended
beyond the optical disk and produces gas-to-dust ratios comparable to
average Milky Way values (Alton et al. 1998, Davies et al. 1999). Cool
dust is also present in elliptical galaxies (Haas 1998), complementing
the IRAS detections of relatively warm dust, T$\sim$25--30~K, in these
objects (e.g., Goudfrooij \& de Jong 1995). Conversely, the ISO
emission of luminous and ultraluminous FIR galaxies, like Arp244,
NGC6240, Arp220, and M82, are fit by a modified Planck function with
color temperature $\sim$30--50~K, for emissivity index $\epsilon$=1
(Klaas et al. 1997, Colbert et al. 1999).

Our investigation concentrates on actively star-forming galaxies,
because of the key role they play in interpreting high redshift
galaxies. Some of the observational characteristics of high redshift,
z$\sim$3 and z$\sim$4, Lyman-break galaxies (Steidel et al. 1996,
1999, Giavalisco et al. 1996, 1999) are similar to those of the
central regions of local starburst galaxies. Such characteristics
include the star formation rates per unit area, the shape of the
stellar continuum and the intensity and shape of the absorption
features in the restframe UV spectra, and the ionizing/UV-continuum
photon ratios (Steidel et al. 1996, Pettini et al. 1998, Meurer,
Heckman \& Calzetti 1999). Still open is the issue of the dust opacity
in these UV-bright, star-forming systems. A new challenge to the
common wisdom that there is little dust opacity in the young Universe
has come from the discovery of FIR-bright galaxies at
intermediate/high redshift with SCUBA (Hughes et al. 1998, Eales et
al. 1998, Blain et al. 1999, Lilly et al. 1999a, Barger et
al. 1999). The necessity to link the two populations of high-redshift
galaxies, UV-bright and FIR-bright, has clearly brought new urgency to
the problem of understanding the opacity and dust emission
characteristics of star-forming galaxies at all redshifts.

This paper presents ISO photometric observations of eight nearby
star-forming galaxies in the wavelength range 120--240~$\mu$m, in
order to study the galaxies' dust content in a wavelength range which
is beyond the limits of IRAS. For five of the galaxies, the balance
between the UV--optical--nearIR stellar energy absorbed by the dust  
and the FIR energy emitted by dust is investigated in detail. The
results from our small sample will be placed in the more general
context of dust opacity in star-forming galaxies at low and high
redshift.

\section{Sample Selection, Observational Strategy and Data Reduction}

The original sample of 40 star-forming galaxies from which our ISO
sources were selected was derived from the IUE Atlas by Kinney et
al. (1993), with the following characteristics: the galaxies are
UV-bright and have star formation rates (SFRs) of a few up to a few
tens M$_{\odot}$~yr$^{-1}$; for each galaxy, we have UV, optical and,
in some cases, near-IR SEDs obtained from large-aperture spectroscopy
(Calzetti, Kinney \& Storchi-Bergmann 1994, McQuade, Calzetti \&
Kinney 1995, Storchi-Bergmann, Kinney \& Challis 1995, Calzetti,
Kinney \& Storchi-Bergmann 1996); the dust reddening properties of
these galaxies in the wavelength range 0.12--2.2~$\mu$m are well
characterized (Calzetti, Kinney \& Storchi-Bergmann 1994, Calzetti
1997a). From this original sample, we selected a subsample of 16
objects with positive IRAS detections at 60~$\mu$m and with more than
50\% of both the blue light and the H$\alpha$ emission included within
the aperture of the UV--optical observations. The last requirement
minimizes aperture mismatches between the short (UV, optical and
near-IR) and long (FIR) wavelength observations. Visibility
constraints and observing time limitations reduced the final number of
targets to the eight listed in Table~1.

The eight objects have redshifts z$<$0.03, metallicities in the range
0.1--2~Z$_{\odot}$, intrinsic color excess in the range
E(B$-$V)=0--0.7~mag, and SFRs in the range
0.2--55~M$_{\odot}$~yr$^{-1}$, as derived from the H$\alpha$ emission
for a 0.1--100~M$_{\odot}$ Salpeter IMF.  The galaxies are non
interacting, although some display double luminous peaks in their
nuclei, perhaps a sign that they are caught at some stage of a merging
process. For all eight objects, between 82\% and 96\% of the H$\alpha$
emission is contained within the UV--optical observational aperture,
and, for seven of them, more than 80\% of the B band emission is
included in the aperture; for NGC7673, the B band fraction in the
aperture is 50\%. These values have been derived from the literature
or, for the H$\alpha$ emission of NGC7673, from new imaging data
(C. Conselice, 1999, private communication). For each galaxy, Table~1
reports the general properties: distance, average angular size,
intrinsic optical color excess, SFR, and UV spectral slope $\beta$
(defined as F($\lambda$)$\propto\lambda^{\beta}$, see Calzetti et
al. 1994). Table~2 lists the photometry at 12~$\mu$m, 25~$\mu$m,
60~$\mu$m, and 100~$\mu$m from IRAS, as well as the photometry at
150~$\mu$m and 205~$\mu$m from our ISO observations.

The ISO data were obtained with the ISOPHOT instrument and the C200
camera with the 135 and the 200 filters, centered at $\sim$150~$\mu$m
and $\sim$205~$\mu$m, respectively (ISOPHOT Observer's Manual
1994). Sparse maps were obtained in each filter, with one on-source
and two off-source exposures per target. The strategy of the double
off-source observation was adopted to average out the uneven FIR
background, as inferred from the IRAS maps. The background positions were
located $\sim$8--10 arcmin away from the target (equivalent to about 3
times the C200 field of view), on blank sky positions.  The two
off-source exposures were always taken before the on-source exposure,
to avoid transients. Typical exposure times ranged between 32~s and
128~s.

The data reduction and calibration were performed with the
PIA\footnote{PIA is a joint development by the ESA Astrophysics
Division and the ISOPHOT Consortium.} package at the Infrared
Processing and Analysis Center (IPAC), following the standard
procedure for point sources. This is appropriate for our galaxies
which are much smaller than the field of view of the C200 Camera
($\sim$3~arcmin, see Table~1). Data reduction and calibration
included correction for the non-linear response of the detectors,
read-out deglitching, linear fitting of the signal ramps, resetting of
all ramp slopes to a common time interval, flagging of the deviant
slopes, dark current subtraction, average of all the ramp slopes for
each pixel, and power calibration against the reference lamp. After
extraction of the flux densities, the two off-source exposures were
averaged and subtracted from the on-source exposure.  Finally, the
source signal in the four pixels of the C200 was added together and
divided by the fraction of the Point Spread Function included within
the C200 field of view. ISOPHOT absolute calibrations have systematic
uncertainties of about 25--30\%, which are the dominant source of
uncertainty in our data for well detected galaxies. The 205~$\mu$m
measurements of NGC5860 and IC1586 are exceptions: in each case the
two backgrounds yielded very different values, and this difference
dominates the uncertainty attributed to the detection. For faint
detections or upper limits, statistical uncertainties
dominate. Table~2 reports the largest among the various sources of
uncertainty for the F(150) and F(205) flux densities. Color
corrections due to dust temperature variations are discussed in
section~3.1 and are not included in the flux density values of
Table~2.

Long wavelength ISO measurements for one of our galaxies, NGC6090,
have already appeared in the literature (Acosta-Pulido et al. 1996). A
recent re-calibration of the published data (J. Acosta-Pulido 1998,
private communication) gives F(150)=8.26~Jy and F(205)=7.16~Jy. While
Acosta-Pulido et al.'s and our determinations of the F(150) flux agree
well within 1~$\sigma$, the two F(205) flux density determinations
differ by 3~$\sigma$. For the data of Acosta-Pulido et al., there is
no independent measurement of the background, which may be the source
of the discrepancy, because the long-wavelength PSFs fill the C200 field
of view. For reference, the average background values we measure
around NGC6090 are 8.9$\pm$0.4~MJy~sr$^{-1}$ at 150~$\mu$m and
4.1$\pm$0.4~MJy~sr$^{-1}$ at 205~$\mu$m, where the quoted
uncertainties are statistical only.

The last two columns of Table~2 list for each galaxy the integrated
FIR fluxes in the range 40--120~$\mu$m and 40--240~$\mu$m,
respectively. The flux in the range 40--120~$\mu$m is from the IRAS
60~$\mu$m and 100~$\mu$m measurements, using the formula of
Lonsdale--Persson \& Helou (1987); the flux in the range
120--240~$\mu$m is derived from our ISO photometry, after
approximating the filter bandpasses to square-equivalent filters of
60~$\mu$m width centered at 150 and $\sim$205~$\mu$m, respectively
(ISOPHOT Observer's Manual 1994), and after including the color
corrections discussed in section~3.1. The observed SEDs of the eight
galaxies are shown in Figure~1. The following sections
concentrate on the five galaxies with positive ISOPHOT detections at
both 150~$\mu$m and 205~$\mu$m and contain no further discussion on
the three galaxies undetected at 205~$\mu$m.

\section{Analysis and Results}

\subsection{Modelling the Emission from Large Dust Grains}

The combination of IRAS and ISO photometry directly measures the FIR
emission in the wavelength region 8--240~$\mu$m.  In this section, we
model the long wavelength ($\lambda>$40~$\mu$m) FIR emission with the
goal of obtaining a reliable extrapolation of the dust emission beyond
240~$\mu$m, and an estimate of the contribution from cool dust
(T$<$25--30~K) to the total FIR luminosity. For this purpose, we
assume that the emission longward of 40$\mu$m is fully accounted for
by large grains in thermal equilibrium with their environment.
Modelling of the emission below $\sim$40~$\mu$m is not attempted
because of its complex characteristics (see the Introduction).

A range of parameters is explored for fitting the four data points
with $\lambda>$40~$\mu$m: the emission is modelled with a single and
with a combination of two modified Planck functions, and the dust
emissivity index is allowed to have values $\epsilon$=1 and
$\epsilon$=2. This approach is not new (e.g., Chini et al. 1986b,
Lonsdale-Persson \& Helou 1987; see also Rowan-Robinson \& Crawford
1989 and some of the references listed in the Introduction) and does
not account in detail for the continuum of temperatures and physical
conditions of the dust in galaxies, but is nevertheless adequate for
describing their long wavelength FIR emission. Through $\chi^2$
minimization, the best fit for the dust temperature(s), and, in the
case of two modified Planck function, for the fraction of thermal FIR
emission contributed by each dust component are searched. The problem
is well constrained, as the number of free parameters is less or equal
the number of independent data points, i.e. there are two and four
free parameters in the single and the two modified Planck function
models, respectively. One additional constraint to the fit is given by
the F(25) flux density: the F(25)/F(60) ratios of our galaxies range
between 0.11 and 0.25, implying that between $\sim$25\% and $\sim$75\%
of the F(25) flux is contributed by large grains (Boulanger et
al. 1988). In the single blackbody model, the temperature is allowed
to vary between 10~K and 80~K. For the two-blackbody model, the warm
dust temperature may vary between 30 and 80~K, while the cool dust
temperature varies between 10 and 30~K. For the temperature range of
interest, color corrections to the flux densities at 60, 100, 150, and
205~$\mu$m are about 1.02, 1.03, 1.16, and 1.09, respectively, and
have been applied to the observed values. The robustness of our
fitting routine has been tested against other authors' results for the
dust emission of FIR-bright galaxies (Klaas et al. 1997, Colbert et
al. 1999); under the same assumptions for the dust emission model, we
reproduce the published best fit temperatures within 3~K.

For three of the galaxies in our sample (NGC6090, NGC7673, IC1586),
the combination of two modified Planck functions yields the best fit
to the observed FIR SEDs, with a reduced $\chi^2$ in the range
1.1--2.2. The warm dust component contributes between 50\% and 75\% of
the F(25) flux. For NGC5860 and Tol1924$-$416, a single modified
Planck function is the best model to describe the long wavelength
emission from the galaxies. In all five galaxies, a dust emissivity
index $\epsilon$=2 is a better fit to the FIR SED than $\epsilon=1$,
with a factor $\sim$1.2--1.5 improvement in the reduced $\chi^2$. The
first five panels of Figure~1 show the best fit curves for the five
galaxies and, for comparison, a fit to the data using a single
blackbody with dust emissivity index $\epsilon$=1 and temperature in
the range 38--61~K (depending on the galaxy). Table~3 lists the
temperatures of the warm and cool components, the fraction of the
thermal FIR luminosity contributed by each component, and the
associated dust masses (see next section). Table~4 lists the thermal
FIR flux in the range 1--1000~$\mu$m contributed by the warm$+$cool
dust (F$_{th}$(1--1000)) and the total FIR flux contributed by both
large and small grains at any temperature (F(1--1000)).
F$_{th}$(1--1000) is calculated by adding the best fit model fluxes in
the ranges 1--40~$\mu$m and 240--1000~$\mu$m to the {\it observed}
flux in the range 40--240~$\mu$m (Table~2). In F(1--1000), the best
fit model fluxes in the range 8--40~$\mu$m are replaced with the
observational data. Below 8~$\mu$m, we estimate that the contribution
of the hot dust and small~grain emission to F(1--1000) is less than
$\approx$5--10\% of the total and is neglected. Table~4 also gives the
ratio of the F(240--1000) flux to the thermal FIR emission. In all
five cases considered, such contribution is no more than a few
percent, suggesting that the IRAS$+$ISO measurements detect the bulk
of the FIR emission from starburst galaxies. This neglects the
potential presence of dust with temperatures T$\lesssim$15~, but the
flux contribution of these additional components to the FIR emission
is not expected to be large.

In NGC6090, NGC7673, IC1586, the contribution to the FIR flux of dust
at T$\sim$20--23~K is between 1/3 and 2/3 of the FIR emission, a
non-negligible fraction of the total; the rest is contributed by warm
dust at temperatures in the range T$\sim$40--55~K.  For the two other
galaxies, NGC5860 and Tol1924$-$416, the best fit models give
temperatures T$\sim$32~K and T$\sim$50~K, respectively. NGC5860 does
not have a cool component (according to our definition), but its
`warm' dust has a fairly low temperature, i.e., it is the coolest
among the galaxies in our sample. Conversely, Tol1924$-$416 is the
only object with a high-temperature warm dust component without a
substantial cool dust component. Even taking into account the
1-$\sigma$ uncertainties of the ISO data, very little, if any, FIR
contribution from dust cooler than $\sim$50~K can be accomodated for
this galaxy, unless the dust is colder than $\sim$15~K (which is
beyond the sensitivity range of our ISO observations).

\subsection{Dust Masses and Gas-to-Dust Ratios}

Dust masses can be estimated from long wavelength FIR fluxes using
fairly standard recipes (Hildebrand 1983, Lonsdale-Persson \& Helou
1987). We adopt the formulation of Young et al. (1989):
\begin{equation}
M_{dust}= C\, S_{100}\, D^2\, [exp(143.88/T_{dust}) -1],
\end{equation}
where $M_{dust}$ is in solar masses, $S_{100}$ is the observed flux at
100~$\mu$m in Jy, $D$ is the galaxy distance in Mpc and the expression
in square brackets is the temperature-dependent part of the
blackbody emission. The constant C is a combination of fundamental
constants and of parameters which depend on the physics of the dust
grains, with value $\sim$1.2~M$_{\odot}$~Jy$^{-1}$~Mpc$^{-2}$ for
$\epsilon$=2 and $\sim$6~M$_{\odot}$~Jy$^{-1}$~Mpc$^{-2}$ for
$\epsilon$=1 (Hildebrand 1983, Draine \& Lee 1984). Table~3 lists
the dust masses associated with the warm and cool dust components from 
the previous section.

In NGC6090, NGC7673 and IC1586, the cool dust mass is between 40 and
150 times the warm dust mass, and the total mass in dust is between
10$^7$ and 10$^8$~M$_{\odot}$. The corresponding HI-to-dust ratios,
corrected for 10\%~ He abundance, are listed in the last column of
Table~3. For reference, typical HI-to-dust ratios are 100-150 for the
Milky Way (e.g., Sodroski et al. 1997). If the H$_2$ gas mass is
included in the balance (Young et al. 1995), the gas-to-dust ratio of
NGC6090 is $\approx$1.7 times larger than the Milky Way average
(Bohlin, Savage \& Drake 1978). The two galaxies have similar
metallicity, thus they are expected to have approximately similar
gas-to-dust ratios; this is what is observed, within the uncertainties
of the various estimates. NGC7673 and IC1586 have about half
the metal content of the Milky Way (e.g., Calzetti et al. 1994),
implying that their gas-to-dust ratios are expected to be a factor 2
larger than the Galaxy value, for similar dust depletion
patterns. Using upper limits on the CO luminosity of NGC7673 (Gordon,
Heidmann \& Epstein 1982) to estimate its H$_2$ gas content, we find
(gas/dust)$_{NGC7673}\sim$2(gas/dust)$_{MW}$, as expected from the
metallicity differences. For IC1586, there are no estimates on the
H$_2$ mass in the literature; we find
(HI/dust)$_{IC1586}\sim$4--5(HI/dust)$_{MW}$, which is about a factor
2 larger than the expectations, after taking into account metallicity
differences with our Galaxy. This is still an acceptable agreement
within our uncertainties, although we cannot exclude that a
non-negligible mass of dust colder than T$\sim$15~K may exist in
IC1586.

The total dust mass derived from the two-blackbody fit and
$\epsilon$=2 emissivity is within a factor 4 of the dust mass derived
from a single blackbody model with $\epsilon$=1 emissivity (column~9
of Table~3). Therefore, either model produces reasonably consistent
masses for dust emitting in the temperature range 20--60~K.

\subsection{The Energy Balance}

This section presents a comparison between the measured amount of
stellar energy absorbed by dust, represented by the observed FIR
emission, and a predicted value of the dust-absorbed stellar energy 
derived from the UV--to-nearIR SED of each galaxy in our sample. The
goal is to establish whether, and within which accuracy, extant
methods for UV--to--nearIR dust reddening correction (the `starburst
reddening' curve of Calzetti et al. 1994, see also Calzetti 1997a) can
also be used to derive the total dust optical depth in UV-bright,
starburst galaxies. For this purpose, the predictions on the
dust-absorbed stellar light will be derived by applying the starburst
reddening curve and the measured color excess values E(B$-$V)
(Table~1) to the observed UV--to--nearIR SEDs of the galaxies.

Four of the five galaxies have data in the wavelength region
0.12--2.2~$\mu$m and the fifth, Tol1924$-$416, has data in the region
0.12--1.0~$\mu$m from the literature (Kinney et al. 1993, McQuade,
Calzetti \& Kinney 1995, Storchi-Bergmann, Kinney \& Challis 1995, and
Calzetti, Kinney \& Storchi-Bergmann 1996). As discussed in Section~2,
the galaxies in the present sample have been chosen to minimize the
aperture mismatch between the UV--optical observations (effective area
of $\sim$170~arcsec$^2$) and the FIR photometry (with effective areas
of a few arcminutes$^2$). In addition, we have an estimate of the
fraction of light falling outside the UV--optical--nearIR
observational apertures (fourth column of Table~1), which we use to
correct for the residual effect of aperture mismatch. All UV-to-nearIR
SEDs have been extrapolated down to the Lyman break at 0.0912~$\mu$m
using a power law fit to the UV spectrum. Table~5 lists the observed
flux, F$_o$, in the waveband 0.09--2.3~$\mu$m for the five galaxies.

The intrinsic shape of the stellar emission, F$_i$($\lambda$), is
recovered using the starburst reddening curve
$k^{\prime}$($\lambda$)=A$^{\prime}$($\lambda$)/E$_s$(B$-$V), with the
standard formulation (Calzetti et al. 1994, Calzetti 1997b):
\begin{equation}
F_i(\lambda) = F_o(\lambda)\ 10^{0.4 E_s(B-V)\ k'(\lambda)},
\end{equation}
with F$_i(\lambda)$ and F$_o(\lambda)$ the intrinsic and observed
stellar continuum flux densities, respectively; the color excess of
the stellar continuum E$_s$(B$-$V) is linked to the color excess
derived from the nebular gas emission lines E(B$-$V) (Table~1) via:
\begin{equation}
E_s(B-V) = (0.44\pm 0.03) E(B-V)
\end{equation}
(Calzetti 1997b). The expression of $k^{\prime}$($\lambda$) is:
\begin{eqnarray}
k'(\lambda) &=& 2.659\, (-1.857 + 1.040/\lambda) + R^{\prime}_V \ \ \ \ \ \ \ \ \ \ \ \ 0.63\ \mu m \le \lambda \le 2.20\ \mu m \nonumber \\
           &=& 2.659\, (-2.156 + 1.509/\lambda - 0.198/\lambda^2 + 0.011/\lambda^3) + R^{\prime}_V \nonumber \\
           & &\ \ \ \ \ \ \ \ \ \ \ \ \ \ \ \ \ \ \ \ \ \ \ \ \ \ \ \ \ \ \ \ \ \ \ \ \ \ \ \ \ \ \ \ \ \ \ \ \ \ \ \ 0.12\ \mu m \le \lambda < 0.63\ \mu m.
\end{eqnarray}
The amount of UV-to-nearIR stellar continuum light absorbed by dust is
the difference F$_i-$F$_o$ between the intrinsic and the observed
emission. If the entire scattering area of the galaxy is contained
within the observational aperture, the reddening curve represents
mainly an `absorption' curve. Such an assumption is appropriate for
our observational configuration, where large apertures are employed,
and a large fraction of the light from the galaxy and practically all
the light from the starburst are observed. The energy balance between
UV--to--nearIR and FIR must include the fraction of ionizing photons
($\lambda<$0.0912~$\mu$m) absorbed by dust, either directly or after
they have been re-emitted by hydrogen as Ly-$\alpha$ photons; this
fraction is about 2/3 of the total, albeit with a large uncertainty
(see references by Calzetti et al. 1995). An estimate of the ionizing
energy is obtained from the reddening-corrected H$\alpha$ flux, using
the formula of Leitherer \& Heckman (1995). The predicted
dust-absorbed stellar flux is then given by: $\Delta F(0-2.3) = F_i -
F_o + 0.67 F_{ion}$.  Table~5 lists both the ionizing flux and the
`intrinsic' stellar continuum flux F$_i$, integrated from 0.09~$\mu$m
to 2.3~$\mu$m. The largest uncertainty in the derivation of
$\Delta$F(0$-$2.3) is in the R$_V^{\prime}$=A$^{\prime}$(V)/E$_s$(B$-$V)
parameter of the starburst reddening curve. If we were dealing with
standard extinction curves (rather than a reddening curve for
galaxies), R$_V^{\prime}$=R$_V$ would be the total extinction at V,
which has value R$_V$=3.1 for the Galactic diffuse interstellar
medium. For starburst galaxies the exact meaning of R$_V^{\prime}$ is
less straightforward, since extinction, scattering, and the geometrical
distribution of the dust relative to the emitters are all folded
together. However, R$_V^{\prime}$ can still be described as an
effective total obscuration at V (see Calzetti 1997b, Meurer et
al. 1999). 

Using the value R$_V^{\prime}$=(4.88$\pm$0.98) (Calzetti 1997b) to
calculate $\Delta$F(0$-$2.3) yields the predicted-to-observed
dust-absorbed light ratios $\Delta$F(0$-$2.3)/F(1$-$1000) listed in
Table~5. The ratios vary between 0.6 and 2.1, with average
$<\Delta$F(0$-$2.3)/F(1$-$1000)$>$=1.18 (logarithmic average), for the
four galaxies with non-negligible reddening at optical wavelengths,
i.e. NGC6090, NGC7673, NGC5860, IC1586. In the ideal case that the
reddening curve fully recovers the light lost to dust, the ratio
$\Delta$F(0-2.3)/F(1-1000)=1, because of energy conservation. This
means that, on average, the starburst reddening curve with
$R_V^{\prime}$=4.88 overpredicts the dust FIR flux by about 18\%. In
order to decrease the predicted FIR flux by this excess, the
effective obscuration at V must be decreased down to:
\begin{equation}
R_V^{\prime} = 4.05 \pm 0.80
\end{equation}
Although this value $R_V^{\prime}$ is derived from the average of only
4 galaxies, we will show in section~4.2 that it has a more general
validity, as it can be applied to large samples of star-forming
galaxies. It should also be noted that equation~5 is only 1~$\sigma$
away from the original value derived by Calzetti (1997b).

The dispersion around the mean for $<\Delta$F(0$-$2.3)/F(1$-$1000)$>$
is about a factor 2 (column~5 of Table~5). Reasons for such dispersion
can be varied. In cases where the application of the reddening curve
underestimates the FIR flux, like in NGC6090, some regions of the
central starburst may be heavily enshrouded in dust and they give
little, if any, contribution to the observed UV-to-nearIR flux. In
cases where the reddening curve overpredicts the FIR emission, as for
NGC5860, variations in the geometry of the stellar populations and the
dust (e.g., the entire scattering region is not included in the
observational aperture) can play a role. A factor of 2 dispersion in
the predicted FIR flux of starburst galaxies is, however, remarkably
small, given that a single recipe for dust obscuration corrections is
being used. Thus, within the limitations of the small sample available
to us, the curve recovers the total dust optical depth of UV-bright,
starburst galaxies and can be appropriately called an `obscuration
curve'.
 
\section{Discussion}

\subsection{Far-Infrared Emission from Local Starburst Galaxies.}

Dust emission at wavelengths beyond 120~$\mu$m (the IRAS limit)
represents a non negligible fraction of the FIR flux even in `warm'
galaxies like starbursts. Table~4 lists the bolometric corrections
between the total FIR dust emission in the wavelength window
1--1000~$\mu$m and the 40--120~$\mu$m IRAS measurements for the five
galaxies in our sample, which has average:
\begin{equation}
BC_d = {F(1-1000)\over F(40-120)}= 1.75\pm 0.25.
\end{equation} 
The dust bolometric corrections lower to values between 1.4 and 1.8,
with an average of 1.6, if only the thermal dust emission is included
in the balance. Rigopoulou et al. (1996) measured FIR SEDs for a
sample of ultraluminous IRAS galaxies using sub-mm observations. Dust
bolometric corrections derived from seven galaxies in this sample by
Meurer et al. (1999) give BC$_d$(40--1000)=1.4$\pm$0.2 in the
wavelength region 40--1000~$\mu$m, coincident with the median value we
obtain in the same wavelength range, i.e.,
BC$_d$(40--1000)=1.38$\pm$0.22.

Buat \& Burgarella (1998) derived dust bolometric corrections for
actively star-forming galaxies by constraining the dust emission with
IRAS data and the mm observations of Andreani \& Franceschini (1996).
With their calibration, the total-to-IRAS correction for five of the
galaxies in Table~1 (NGC6090, NGC7673, NGC5860, IC1586, and Mrk66) is
around 1.23, and it is BC$_d<$1 for Tol1924$-$416. The corrections
are on average $\sim$43\% smaller than ours, and the discrepancy
increases for increasing deviation of the galaxies' FIR emission from
the single blackbody approximation. This result suggests that
mm-wavelength observations alone, even when coupled with IRAS data,
are inadequate to constrain the long wavelength FIR
emission. Observations in the mm regime fall in the Rayleigh-Jeans
tail for all dust components warmer than a few K and are difficult to
use for deriving the actual emission in the sub-mm range (see, e.g.,
Chini et al. 1986b, Roche \& Chandler 1993).

For the three galaxies in our sample whose thermal FIR SED is modelled
by a combination of two modified Planck functions (Table~3), the cool
dust component represents between 30\% and 60\% of the thermal FIR
emission. This component is very likely heated by a combination of the
central starburst and the ISRF. The comparison of NGC6090 with NGC7673
provides support to this statement. The SFRs of the two starbursts, as
derived from their extinction corrected H$\alpha$ luminosities, differ
by a factor $\sim$11 (Table~1, after including aperture corrections), while
their host galaxies differ by less than a factor 3 in R-band
brightness, with NGC6090 being the brighter and more active of the
two. If the cool dust were heated mostly by the ISRF, the fraction of
FIR luminosity due to warm dust should be larger in NGC6090 than in
NGC7673 (Young et al. 1989), i.e.
(f$_w$/f$_c$)$_{NGC6090}\approx$4(f$_w$/f$_c$)$_{NGC7673}$. Instead,
(f$_w$/f$_c$)$_{NGC6090}\sim$(f$_w$/f$_c$)$_{NGC7673}$,
implying that most of the cool dust heating in NGC6090 is due to the
starburst itself. A similar conclusion is reached by comparing NGC6090
with IC1586.  The dust temperature decreases for increasing distance
of the dust grains from the heating source, for increasing dust column
density (optical depth), and for increasing mean wavelength of the
source SED (Panagia 1978). Because of the exponential dependence of
$T$ on the medium's optical depth, large dust opacities reduce the
temperature even in the presence of hard radiation from massive
stars. Optically thick regions introduce a spread towards low
temperatures of the emission spectrum because of dust self-shielding
and shielding of dust grains by H$_2$ (e.g. Natta et al. 1981, Mathis
Mezger \& Panagia 1983 for the GMCs in the Milky Way, Wall et al. 1996
for COBE observations of the Orion Constellation). Luminous FIR
galaxies, like NGC6090, are candidates for an opaque medium at most
wavelengths (Lisenfeld, Isaak \& Hills 1999): in the environment of
luminous and ultraluminous FIR galaxies, the gas densities are large
and can reach about 10 times the density of GMCs (Downes \& Solomon
1998); dust column densities are, thus, likely to be large.

Effects of dust optical depth explain qualitatively why Tol1924$-$416
does not have a cool dust component and its dust is generally warm,
with T$\sim$50~K.  Tol1924$-$416 is the most metal-poor and dust-poor
galaxy in our sample of five, with a metallicity about 5--10 times
lower than the other four and a very small optical dust reddening
(column~5 of Table~1). Comparably low dust opacities are expected if
the gas column densities are similar among the five galaxies.  Dust
self-shielding is less effective in Tol1924$-$416 than in the other
galaxies, and the radiation from the massive stars will heat the dust
to relatively high temperatures in the entire starburst region.

Optically thick regions are not the only cause for the presence of a
cool dust component; the non-ionizing stars produced in the central
star-forming region can also heat the dust to relatively low color
temperatures, similar in effect to the ISRF. For typical
star-formation durations of $\sim$50--100~Myr or more, a number of
long-living, non-ionizing stars are accumulated in the starburst site
and become important for the dust heating. Photons from these stars
are less energetic on average than the radiation from massive stars
and can also travel long distances before being intercepted by a dust
grain (Mezger et al. 1982). The dust is then heated to relatively low
color temperatures (e.g., the case of the postburst galaxy NGC4945,
Koornneef 1993). For the typical SFRs of NGC7673, NGC5860, and IC1586
(Table~1), the energy density of the starburst's non-ionizing stars
equals that of the host galaxy diffuse radiation field at a distance
about 0.6~times the radius of the galaxy from the starburst center, in
the assumption that the host galaxy has stellar population
characteristics similar to the Milky Way. Since $<u> \propto
T^{4+\epsilon}$, a factor of a few variation in the energy density
$<u>$ produces only a few percent change in the dust temperature $T$.

Our galaxies are generally classified as `UV-bright'.  However, with
the exception of Tol1924$-$416 which contains relatively little dust,
all the others are affected by dust obscuration in a non-negligible
manner.  The energy emerging at UV-to-nearIR wavelengths is 1/3 or
less of the bolometric luminosity; furthermore, the UV emission below
0.2~$\mu$m represents less than 15\% of the FIR emission (Tables~5 and
6, and Figure~1). A `picket-fence' or 'clumpy' geometry for the dust
can account for the apparent contradiction that such galaxies are at
the same time `UV-bright' and good FIR emitters: most of the
UV-to-nearIR emission will emerge from dust-clear lines of sight,
while the FIR emission emerges from dusty lines of sight (Calzetti
1997a, Gordon, Calzetti \& Witt 1997).

\subsection{The Starburst Obscuration Curve: Application to a Larger 
Sample of Local Galaxies}

The analysis of section~3.3 has shown that the starburst
reddening/obscuration curve of Calzetti et al. (1994) measures the
total dust opacity affecting the UV-to-nearIR SEDs of starburst
galaxies with an accuracy of a factor of about 2 in each individual
case. Up to this point, the validity of such result is limited to the
few galaxies in our sample. In this section, the generality of the
obscuration curve is tested on a much larger sample of `UV-bright',
starburst galaxies.

UV data for a sample of 47 actively star-forming galaxies are drawn
from the Kinney et al. (1993) IUE Atlas and are complemented with IRAS
FIR fluxes. A list of general characteristics of the 47 galaxies are
reported by Meurer et al. (1999), together with the values of the UV
spectral slopes $\beta$, and a discussion on the uncertainties due to
aperture mismatches between IRAS and IUE data. The UV spectral slope
$\beta$, derived in the wavelength range 0.12--0.26~$\mu$m, is a
measure of the dust {\it reddening} affecting the UV stellar continuum
of a starburst galaxy (Calzetti et al. 1994).

Because of the highly incomplete spectral information available,
basically only UV and IRAS FIR data, the energy balance approach of
section~3.3 cannot be applied to the large sample. A more approximate
method is therefore employed here (Meurer et al. 1999): in addition to
a measure of the UV and optical reddening ($\beta$ and E(B$-$V)), an
estimate of the UV total obscuration is derived for each galaxy,
and the two quantities are compared with predictions from the
starburst obscuration curve. In a starburst galaxy most of the
intrinsic stellar light is emitted in the UV, and the FIR-to-UV energy
ratio, F(IR)/F(UV), is an estimator of the global dust optical depth
at UV wavelengths (Xu \& Helou 1996). In the present case,
F(IR)=F(40--120) and F(UV)=[0.16 f(0.16)], with f(0.16) the flux
density at 0.16~$\mu$m. The ratio F(IR)/F(UV) is converted into an
obscuration A($\lambda$) (in magnitudes) at 0.16~$\mu$m via:
\begin{equation}
A(0.16)\simeq2.5\, \log\Bigl({1\over E}{F(IR)\over F(UV)} + 1 \Bigr),\end{equation}
where:
\begin{equation}
E\simeq0.9.
\end{equation}
The constant $E$ is the ratio of two quantities: the bolometric
correction of the UV-to-nearIR stellar light relative to the UV
emission at 0.16~$\mu$m and the dust bolometric correction relative to
the fraction of FIR light detected in the IRAS window. Our value of E
is about 30\% smaller than that of Meurer et al. (1999) for the
following reasons: (1) the dust bolometric correction used in
equation~7 is the one we derive from our ISO observations
(equation~6), which is about 25\% larger than that used by the other
authors; (2) the bolometric correction of the stellar light is derived
under the assumption that a star formation event lasts
$\approx$10$^8$--10$^9$~yr (Calzetti 1997a), rather than $<$10$^8$~yr;
(3) the fraction of stellar light absorbed by dust at each wavelength
is treated semi-analytically, instead of being approximated to its
mean value at UV wavelengths.  Equation~7 has been derived assuming
that the approximation of foreground dust holds for UV-bright
starbursts; various observations support this assumption (Calzetti et
al. 1994, 1996).

A plot of A(0.16) as a function of the UV spectral slope $\beta$ is
shown in Figure~2 for the 47 starburst galaxies. The tight correlation
between A(0.16) and $\beta$ shows that the same amount of dust
responsible for the UV reddening is also responsible for the total UV
opacity (Meurer et al. 1999). The continuous line in Figure~2 is
the correlation between A(0.16) and $\beta$ expected from the
starburst obscuration curve of section~3.3:
\begin{equation}
A(0.16) = 2.31 (\beta - \beta_o) = 2.31\, \beta + 4.85.
\end{equation}
For the no-dust case (A(0.16)=0), an intrinsic UV spectral slope
$\beta_o=-$2.1 has been adopted (Calzetti 1997a), as expected for
continuous star formation over a timescale of 1~Gyr. Once $\beta_o$ is
fixed, there are no other free parameters in equation~9. The excellent
agreement between the locus of the observed points and the predicted
curve suggests that the reddening curve of equation~4, with the
zero-point of equation~5, is a representative {\it obscuration curve}
for UV-bright, star--forming galaxies in general. It recovers fairly
accurately (within $\sim$20\%) the mean dust opacity of a large sample
of such galaxies. This is not the result of a circular argument: the
reddening curve (equation~4) has been derived entirely from
UV--to--nearIR data of starburst galaxies, while its zero-point
(equation~5) has been derived from the FIR data of a small sample of
five galaxies; the data in Figure~2 use an independent set of data:
the FIR emission detected by IRAS for a relatively large number of
star--forming galaxies.

For fixed $\beta$, the scatter about the mean value of A(0.16) is
typically 0.5--0.6~mag (one-sided); thus, a measure of the global
opacity of any individual star--forming galaxy using the starburst
obscuration curve will give a typical uncertainty of a factor
1.6--1.7, not very different from the factor $\sim$2 found in
section~3.3.

For $\beta<-$1.4, the data points of Figure~2 tend to have higher
values of A(0.16) than predicted by equation~9. There are two possible
reasons for this effect. First, the uncertainty of converting the FIR
emission into a UV opacity is large, especially with limited
wavelength coverage. Detailed energy balance analysis should be
performed on the dust-poor galaxies in order to understand whether the
observed underestimate of the opacity is an actual problem or is an
artifact of how A(0.16) is constructed. Second, very blue, dust-poor
galaxies are generally associated with shorter timescales for star
formation (e.g., Calzetti 1997a), implying that, in equation~9,
$\beta_o$ should be changed from $-$2.1 to a more negative value to
properly represent the data points at $\beta\lesssim-$1.4. For
instance, continuous star formation over 0.1~Gyr gives $\beta_o=-$2.35
for solar metallicity (Leitherer \& Heckman 1995). Variations in the
intrinsic stellar population of each galaxy may thus play a small, but not
insignificant role in the details of the agreement between data and
prediction.

For a subset of $\sim$30 starburst galaxies, color excess
measurements for the ionized gas, E(B$-V$), are available from Calzetti et
al. (1994). Figure~3 shows the obscuration A(0.16) as a function of
E(B$-$V). A correlation between the two quantities is present also in
this case, although with a larger scatter than in Figure~2. Formally,
equations~4 and 5 imply a correlation between A(0.16) and E(B$-$V)
expressed as:
\begin{equation}
A(0.16) = 4.39\, E(B-V).
\end{equation}
The predicted correlation marks the lower envelope to the data points
of Figure~3. E(B$-$V) measures the reddening of the ionized gas
(Calzetti et al. 1994), while A(0.16) is a measure of the dust
obscuration affecting the UV stellar continuum; the two are, therefore,
not directly related. The scatter in the relation between A(0.16) and
E(B$-$V) is no worse than the scatter between $\beta$ and E(B$-$V)
found by Calzetti et al. (1994), which is an imprint of the
differential reddening between stellar continuum and ionized gas
(equation~3) and of the variations in the intrinsic stellar population
between galaxies. For fixed E(B$-$V), the scatter about the mean value
of A(0.16) is about 1~mag (one-sided).

One final caveat: the obscuration curve may not be applicable to very
compact, dusty starbursts, where dust optical depths are large (e.g.,
ultraluminous FIR galaxies) and the foreground dust approximation may
not hold. However, the issue is still open.

\subsection{Implications for High-Redshift Galaxies}

As mentioned in the Introduction, the central regions of local
UV-bright, starburst galaxies show some observational characteristics
similar to those of the Lyman-break galaxies at z$>$2.5 (Steidel et
al. 1999). In both cases, the objects are active star--forming systems
with median star formation rates per unit area of
$\approx$1--2~M$_{\odot}$~yr$^{-1}$~kpc$^{-2}$. Another characteristic
they share is the large spread in restframe UV colors (Dickinson
1998). As seen in the previous section, the range of UV spectral
indices of local starbursts is closely linked to the amount of dust
reddening present in the galaxy. Dust reddening has been proposed also
as an explanation for the observed range of UV colors of the
Lyman-break galaxies (Calzetti 1997b, Meurer et al. 1997, 1999,
Pettini et al. 1998, Steidel et al. 1999).

The median of the distribution of observed UV colors corresponds to a
color excess E$_s$(B$-$V)=0.15 for the stellar continuum (Steidel et
al. 1999, using the normalization $R^{\prime}$(V)=4.88 of Calzetti
1997b). The new normalization derived here, $R^{\prime}$(V)=4.05,
implies a negligible change to the color excess, E$_s$(B$-$V)=0.16,
mainly because the new zero-point affects $k^{\prime}$($\lambda$) by
less than 8\% at $\lambda\lesssim$0.17~$\mu$m. Thus, the intrinsic UV
flux at 0.15--0.17~$\mu$m is on average 4.1--4.6 times larger than
what observed. Dust absorbs about 75\%--80\% of the UV light in
Lyman-break galaxies at z$\approx$3, a figure compatible with the
maximum value of 82\%--85\% inferred from the cosmological evolution
of the dust content of galaxies (Calzetti \& Heckman 1999, Pei, Fall
\& Hauser 1999).

The extended wavelength coverage of our ISO sample can be used to
predict the characteristics of the FIR emission from high-redshift,
UV-bright, star-forming galaxies. Table~6 lists for our sample
galaxies the bolometric luminosity and the energy emitted at a few
selected wavelengths normalized to the thermal FIR emission. As
already remarked, the bulk of the energy, more than 2/3 (Table~5),
emerges at FIR wavelengths, with the exception of Tol1924$-$416 which
is the dust-poor galaxy in this sample.  Table~7 lists selected
observational properties of the same galaxies placed at redshifts z=1
and z=3, respectively, in a Universe with $\Omega_M$=1,
$\Omega_{\Lambda}$=0 and H$_o$=50~km~s$^{-1}$~Mpc$^{-1}$.  At z=3, the
five galaxies would have observed magnitudes R$\sim$25.5--27
(restframe $\lambda\sim$0.17~$\mu$m), about 1.0--2.5
magnitudes lower than the $\sim$R$^*$ value determined by Steidel et
al. (1999) for the Lyman-break galaxies; the expected flux densities
at restframe 100, 150, and 205~$\mu$m, which correspond to the
observer's frame 400, 600, and 820~$\mu$m, are generally below the
detection limits reached by the deepest sub-mm surveys ($\sim$2~mJy
for a 5~$\sigma$ detection with SCUBA at the JCMT, Hughes et
al. 1998). A similar conclusion applies to the z=1 case: our galaxies
placed at that redshift would have fairly faint B magnitudes
(restframe 0.22~$\mu$m), and equally faint FIR flux
densities. Tables~8 and 9 give the predicted FIR flux densities and
corresponding bolometric luminosities for selected SEDs (NGC6090,
NGC7673, Tol1924$-$416) covering a range of B or R magnitudes at
redshifts z=1 and z=3, respectively. The predicted FIR fluxes are,
where possible, directly from IRAS or ISO measurements whose
redshifted wavelengths are the closest to recent sub-mm surveys 
(e.g., SCUBA: 450~$\mu$m and 850~$\mu$m). Otherwise, they are
extrapolated from the model SEDs.

Both Tables~8 and 9 show that only very bright galaxies,
L$_{bol}\gtrsim$2--3$\times$10$^{12}$~L$_{\odot}$, are detectable with
the current sensitivity limits of SCUBA, for the range of SEDs
considered here (see, e.g., the review by Lilly et al. 1999b). These
sources are similar to or brighter than the typical local
ultraluminous FIR galaxy, like Arp220. The ability to detect the dust
FIR emission from a high-redshift galaxy depends, however, not only on
the intrinsic luminosity of the source, but also on the details of the
dust SED. There is a factor $\sim$5--10 difference between
Tol1924$-$416 and the other four galaxies in the amount of energy
coming out at restframe 205~$\mu$m relative to the total thermal FIR
emission (see last column of Table~6). For comparison, the
205~$\mu$m-to-total FIR energy in Arp220 is also about 2.5--5 times
larger than in Tol1924$-$416 (from the data of Klaas et al. 1997). The
difference is mainly driven by the absence of a cool dust component in
Tol1924$-$416. As discussed in section~4.1, there is very little dust
emission with temperature T$\lesssim$50~K in Tol1924$-$416,
independent of the chosen dust emissivity, although this result needs 
further confirmation. 

We can speculate whether the long wavelength FIR SEDs of high redshift
galaxies may be expected to be consistent with `purely' warm dust
emission and be similar to the Tol1924$-$416. Low dust temperatures
are the result of high dust column densities and low SFRs, both
contributing to lower the number of UV photons available to each dust
grain. The dust optical depth of Lyman-break galaxies is not known,
but educated guesses can be made with the available
information. High-redshift galaxies are expected to be more metal-poor
than local galaxies ($\approx$1/10~Z$_{\odot}$ at z$\sim$3, e.g., Pei
\& Fall 1995, Calzetti \& Heckman 1999), although preliminary
measurements seem to suggest relatively high abundances in Lyman-break
galaxies (De Mello, Leitherer \& Heckman 1999, Pettini et al.
1999). Low metal abundances do not necessarily imply low dust column
densities: high-redshift galaxies may still have a large reservoir of
gas, if they have not previously undergone a major star formation
phase. Damped Ly-$\alpha$ absorption systems at z$\sim$3 have HI
column densities in the range N(HI)=(2--10)$\times$10$^{21}$~cm$^{-2}$
(Storrie-Lombardi, McMahon \& Irvin 1996), consistent with the total
gas column densities of $\approx$10$^{22}$~cm$^{-2}$ predicted by
models (Calzetti \& Heckman 1999). The result is that the dust optical
depth of high redshift galaxies may be as large as that of local
starbursts like NGC7673. However, Lyman-break galaxies have reddening
corrected SFRs about 10 times larger than NGC7673 itself (Steidel et
al. 1999), implying that about 10 times more UV photons are available
for each dust grain, and that the dust temperature may be as much as
50\% higher than in local starbursts. The possible youth of the
stellar populations in the Lyman-break galaxies (ages
$\lesssim$1--2~Gyr) may also argue in favor of warm dust. In conclusion,
the dust in high-redshift galaxies may be warmer than in NGC6090-like
and NGC7673-like starbursts and, potentially, as warm as in
Tol1924$-$416, implying that the distant galaxies would be
undetectable with current sub-mm instrumentation.

\section{Summary and Conclusions}

The long-wavelength coverage afforded by ISO has been crucial for
characterizing the thermal FIR emission of actively star--forming
galaxies. We observed with ISOPHOT eight starburst galaxies at both
150~$\mu$m and 205~$\mu$m and detected five in both bands. In two of
the detected objects, NGC5860 and Tol1924$-$416, the best fit to the
IRAS$+$ISO dust emission in the wavelength range
$\sim$40--240~$\mu$m is given by a single temperature value with dust
emissivity index $\epsilon$=2. The dust is warm in both galaxies,
T$\sim$32~K in NGC5860 and T$\sim$50~K in Tol1924$-$416, but the
presence of cold dust with T$\lesssim$15~K cannot be excluded in
either case. In the other three objects (NGC6090, NGC7673, IC1586),
the best fit to the thermal dust emission is given by two modified
Planck functions with temperature T$_w\sim$40--55~K for the warm dust
component and T$_c\sim$20--23~K for the cool dust component, both with
dust emissivity index $\epsilon$=2. The warm dust, heated by the
ionizing stars produced in the starburst, contributes between 40\% and
70\% of the thermal FIR emission (e.g., Chini et al. 1986a,
Lonsdale-Persson \& Helou 1987, Young et al. 1989). The rest of the
FIR emission is due to the cool dust, which has an associated mass
around 10$^7$--10$^8$~M$_{\odot}$, i.e., between 40 and 150 times the
warm dust mass. The resulting gas-to-dust ratio values are
compatible, within a factor $\approx$2 with the Milky Way average
(Sodroski et al.  1997) once variations in metallicity are taken into
account. The contribution of the cool dust component is, therefore,
not negligible even in `warm' systems such as starburst galaxies,
although its importance varies by many factors from galaxy to
galaxy. The heating of a fraction of the cool dust component is from
the starburst, as the ISRF alone is not sufficient to account for the
FIR flux associated with the cool dust. Large dust column densities and
non-ionizing photons from the ageing stars are the most likely causes
for cool dust associated with the starburst (e.g., Lisenfeld et al.
1999).

Around 70\% of the bolometric energy is emitted in the FIR, even if
the galaxies in our sample are generally classified as
`UV-bright'. Moreover, the UV emission below 0.2~$\mu$m represents
less than 15\% of the FIR emission (with the exception of
Tol1924$-$416). These estimates are consistent with the amount of
stellar light absorbed by dust at UV-to-nearIR wavelengths as
predicted by the starburst obscuration curve of Calzetti et al. (1994,
Calzetti 1997a). The starburst obscuration curve, applied to SEDs of
star--forming galaxies with a zero-point R$^{\prime}_V\simeq$4.05,
reproduces the observed FIR emission within a factor 2 for individual
objects and within $\sim$20\% when averaged over large galaxy
samples. The improvement in accuracy is due to the averaging of
stellar populations and dust geometry variations in large samples.

The variety of FIR SEDs observed in our sample of local galaxies has
direct implications for the detectability of UV-bright, high-redshift
galaxies at FIR restframe wavelengths. If the bolometric luminosities
of Lyman-break galaxies are around or above
$\sim$3$\times$10$^{12}$~L$_{\odot}$ {\it and} their FIR SEDs resemble
that of NGC6090, they would be detectable with current sub-mm
instrumentation, as the predicted flux at 850~$\mu$m (observer's
frame) is around or greater than 4.4~mJy. However, if the dominant FIR
SED in high-redshift galaxies is like Tol1924$-$416, these galaxies
will be undetected with the current SCUBA sensitivity limits, no
matter how bright the objects intrinsically are. Similar
considerations hold for galaxies at z$\sim$1. There are reasons to
believe that dust in z$\sim$3 galaxies may be hotter than in galaxies
at lower redshifts. Star formation is intense and widespread over the
entire galaxy in the high-redshift systems (Giavalisco et al. 1996,
1999), thus increasing the number of ionizing photons available per
dust grain. In addition, z$\sim$3 galaxies could be fairly unevolved
objects, possibly implying little contribution from aged stellar
populations to the dust heating. Both effects would combine to make
high-redshift (z$\gtrsim$3) galaxies relatively `hot' FIR emitters.

Information on the FIR emission of the Lyman-break galaxies has been
recently collected (Chapman et al. 1999), but very little or only
partial data on the long-wavelength SEDs of these objects or of the
SCUBA sources is currently available (e.g., Lilly et al. 1999, Smail
et al. 1999, and references therein). Measuring dust SEDs is necessary
to determine the dust content and emission characteristics of galaxies
at high redshift and the relationship between the Lyman-break galaxies
and the SCUBA sources. In the low redshift Universe, ISO has only
started to unveil the full extent of the parameter space (e.g., range
of temperatures, masses, spatial distribution of the different dust
components, dependence on metallicity, starburst/host ratio, etc.) of
the dust FIR emission in galaxies. Future FIR missions, such as SIRTF,
together with sub-mm wavelength observations will be instrumental for
investigating the dust content of local galaxies in statistically
meaningful samples.

\acknowledgments

The authors thank Rosemary Wyse for many useful discussions and
comments in different parts and at various stages of this work.
D.C. thanks Nino Panagia for discussions on the FIR opacity mechanisms
in our Galaxy and for a critical reading of the manuscript. The
hospitality of IPAC for two full sessions of data reduction is
acknowledged; in particular D.C. thanks Nanyao Lu for walking
her through the mysteries of the data reduction software
PIA. Literature searches have made use of the NASA/IPAC Extragalactic
Database (NED) which is operated by the Jet Propulsion Laboratory,
California Institute of Technology, under contract with the National
Aeronautics and Space Administration. This work was partially
supported by the NASA ISO grant J-0496/NAG5-3360.

{}


\newpage
\figcaption[figure1.ps]{Spectral energy distributions, from
0.12~$\mu$m to 240~$\mu$m, of the eight galaxies in our sample. UV--optical
spectrophotometry is shown as a continuous line in the wavelength
range 0.12-1.0~$\mu$m, near-IR photometry as empty circles, IRAS data
as filled squares, and our ISO measurements as filled triangles. Upper
limits are shown as downward arrows. No renormalization between the
small UV--optical--nearIR observational apertures
($\sim$20$^{\prime\prime}\times$10$^{\prime\prime}$) and the large
FIR apertures ($\sim$ a few arcminutes) has been performed. For five
of the galaxies, the best fit curve to the dust emission in the
range 40--240~$\mu$m is also shown (solid line). The best fit model 
corresponds to single or two modified Planck functions with the temperatures 
given in Table~3, a dust emissivity $\epsilon$=2, and 50\% 
contribution to the observed F(25) flux density. For comparison, a
model of a single blackbody with $\epsilon$=1 dust emissivity is also shown
(dashed line).}

\figcaption[figure2.ps]{The opacity at 0.16~$\mu$m, A(0.16) plotted as
a function of the UV spectral slope $\beta$ for a sample of 47
starburst galaxies from Kinney et al.'s IUE Atlas. The filled triangles
mark the position of our 5 galaxies; upper limits are
indicated as downward arrows. A representative 1~$\sigma$ error-bar is
given at the bottom-right corner of the diagram. A(0.16) is derived
from the ratio FIR$_{IRAS}$ to UV luminosity. $\beta$ is a measure
of the dust reddening affecting the galaxy. The tight correlation
between the two quantities demonstrates that the same amount of dust
responsible for the reddening is also responsible for the total
obscuration (see Meurer et al. 1999). The continuous line is the trend 
predicted by the starburst obscuration curve k$^{\prime}$($\lambda$)
(equations~4 and 5). The slope of the line is dictated by
the functional shape of k$^{\prime}$($\lambda$), the intercept by the 
UV characteristics of the intrinsic stellar SED (see text).}

\figcaption[figure3.ps]{The opacity at 0.16~$\mu$m, A(0.16), plotted as
a function of the color excess E(B$-$V) for a sample of 30 starburst
galaxies (Calzetti et al.  1994). E(B$-$V) is measured from the
ionized gas emission. Symbols are as in Figure~2. A representative
1~$\sigma$ error-bar is given at the bottom-right corner of the
diagram. The continuous line shows the trend predicted by the
starburst obscuration curve k$^{\prime}$($\lambda$) (equations~4 and
5), after imposing that the opacity of the stellar continuum is zero
(A(0.16)=0) if the reddening of the ionized gas is zero
(E(B$-$V)=0). The predicted line marks the lower envelope of the
correlation between A(0.16) and E(B$-$V).}

\begin{figure}
\figurenum{1}
\plotone{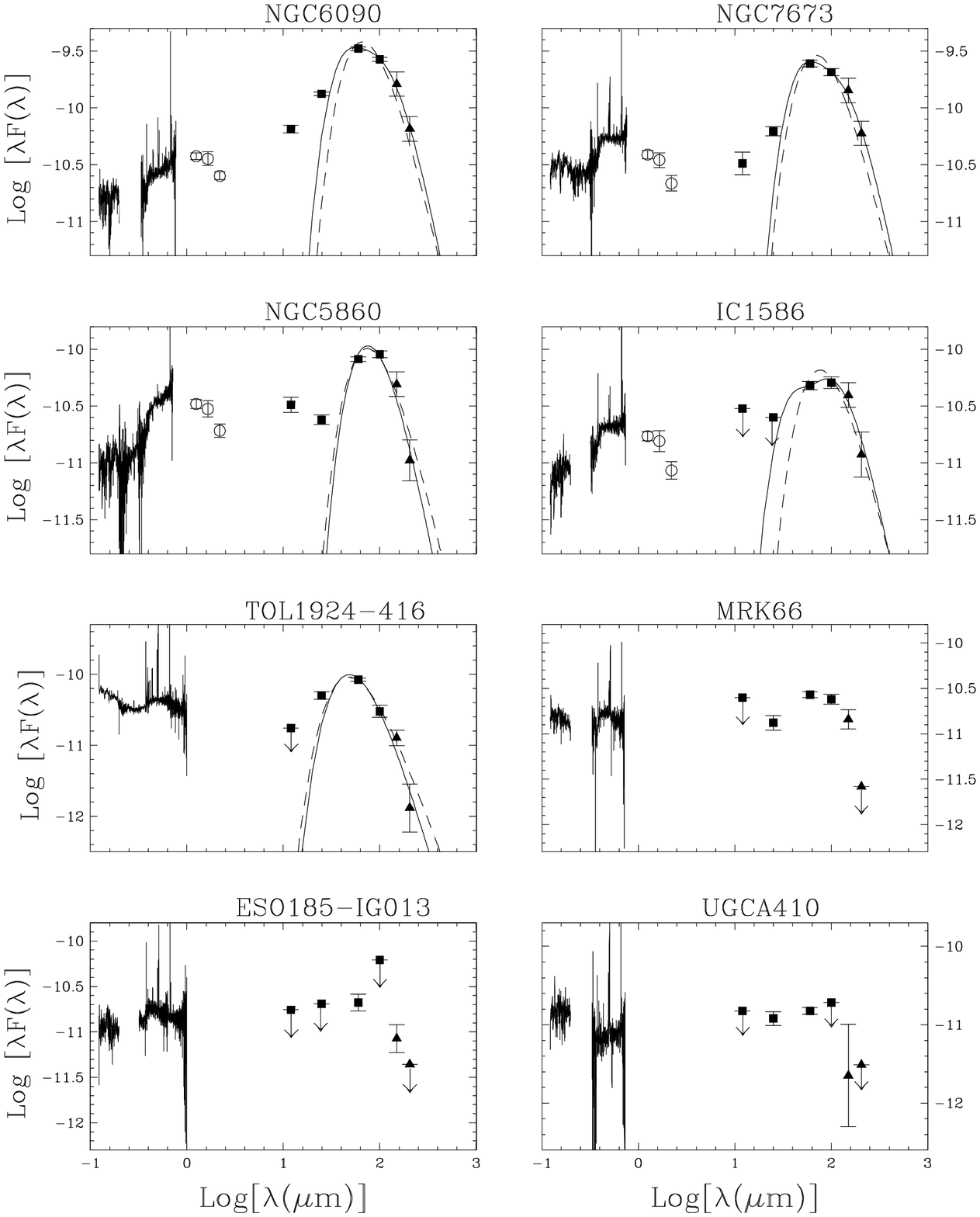}
\figcaption[f1.eps]{}
\end{figure}

\begin{figure}
\figurenum{2}
\plotone{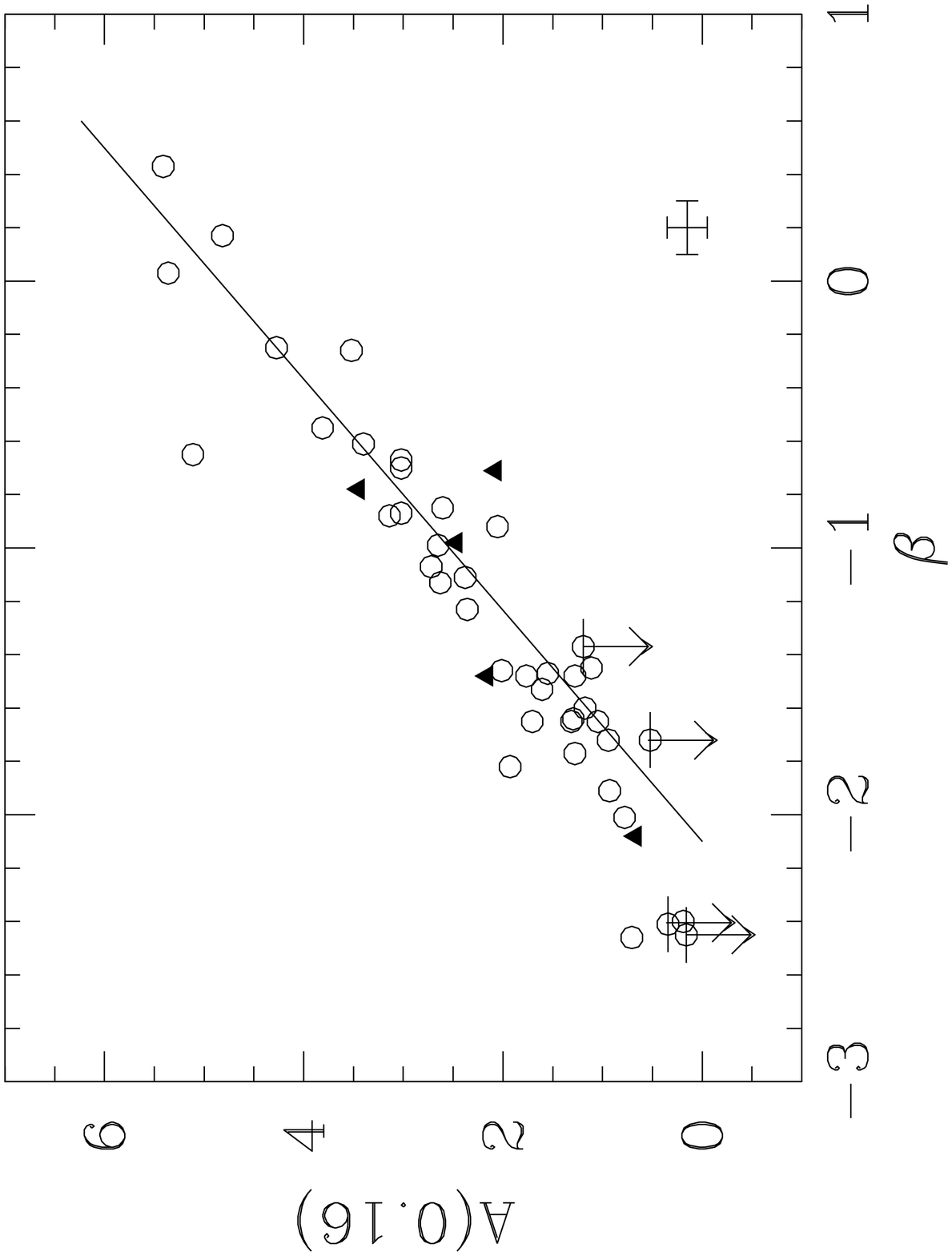}
\figcaption[f2.eps]{}
\end{figure}

\begin{figure}
\figurenum{3}
\plotone{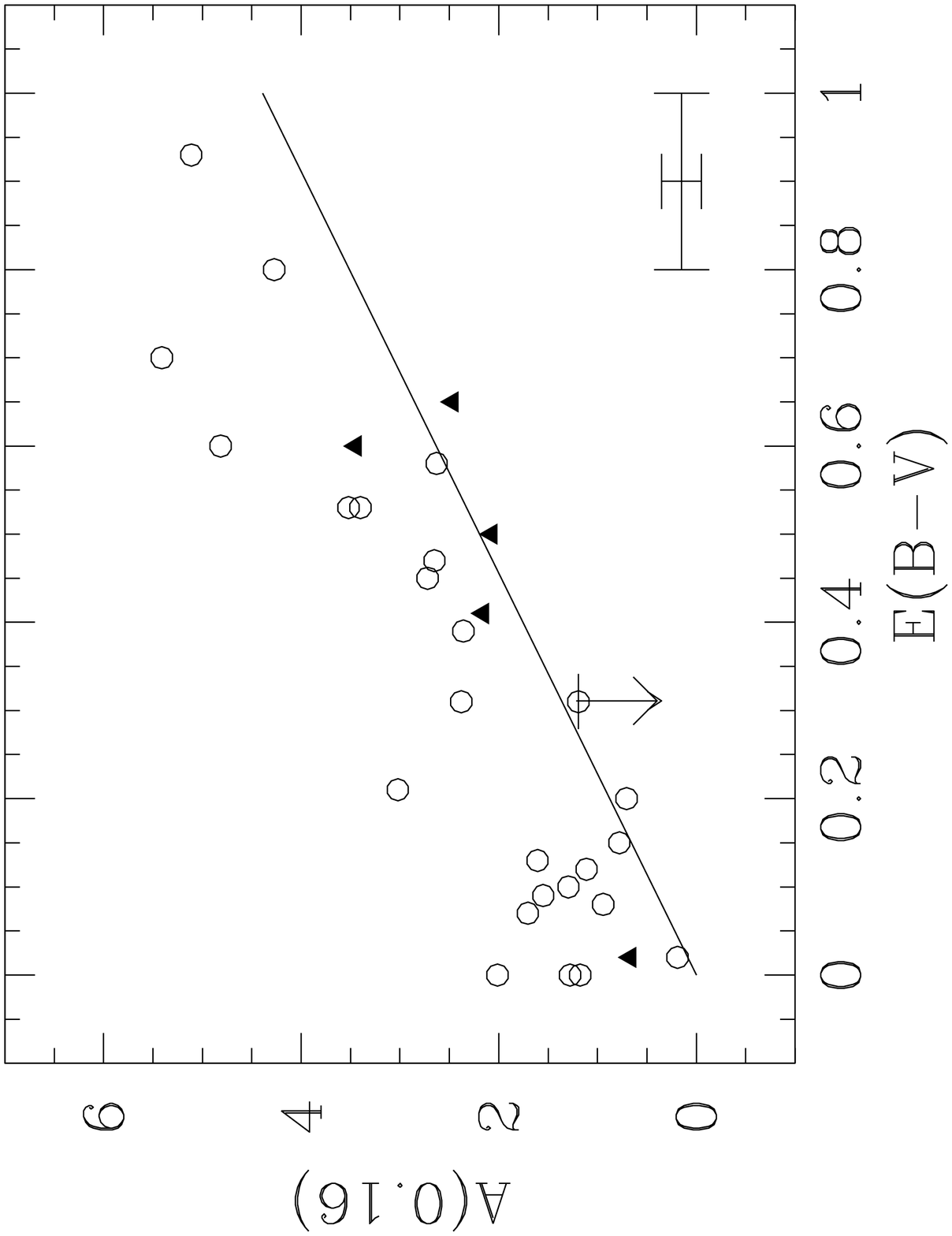}
\figcaption[f3.eps]{}
\end{figure}

\end{document}